\documentclass[floatfix,aps,showpacs,twocolumn,nofootinbib,preprintnumbers]
{revtex4}
\usepackage{graphicx}

\begin{document}

\preprint{DESY 06-089}

\title{Inflation and WMAP three year data: Features have a Future!}
\author{Laura Covi}
\affiliation{Deutsches Elektronen-Synchrotron DESY, Notkestr. 85, 22607
    Hamburg, Germany}
\author{Jan Hamann} 
\affiliation{Deutsches Elektronen-Synchrotron DESY, Notkestr. 85, 22607
    Hamburg, Germany}
\author{Alessandro Melchiorri}
\affiliation{Dipartimento di Fisica and Sezione INFN,
Universita' di Roma ``La Sapienza'', Ple Aldo Moro 2, 00185, Italy}
\author{An\v{z}e Slosar}
\affiliation{Faculty of Mathematics and Physics, University of Ljubljana, Slovenia}
\author{Irene Sorbera}
\affiliation{Dipartimento di Fisica,
Universita' di Roma ``La Sapienza'', Ple Aldo Moro 2, 00185, Italy}

\date{\today}

\begin{abstract}
The new three year WMAP data seem to confirm the presence of
non-standard large scale features in the cosmic microwave anisotropy
power spectrum. While these features may hint at uncorrected
experimental systematics, it is also possible to generate, in a
cosmological way, oscillations on large angular scales by introducing
a sharp step in the inflaton potential. Using current cosmological
data, we derive constraints on the position, magnitude and gradient of
a possible step. We show that a step in the inflaton potential, 
while strongly constrained by current data, is still
allowed and may provide an interesting explanation to the currently
measured deviations from the standard featureless spectrum.
Moreover, we show that inflationary oscillations in the primordial 
power spectrum can significantly bias parameter estimates from 
standard ruler methods involving measurements of baryon oscillations.
\end{abstract}


\maketitle

\section{Introduction}

The recent three year results from the Wilkinson Microwave Anisotropy
Probe (WMAP) satellite \cite{wmap3cosm,wmap3pol,wmap3temp,wmap3beam}
have further confirmed with an extraordinary precision the
inflationary paradigm of structure formation in which primordial
fluctuations are created from quantum fluctuations during an 
early period of superluminal expansion of the universe 
\cite{Starobinsky:1979ty,muk81,bardeen83}.  

Indeed, soon after the WMAP data release, a number of authors 
investigated the possibility to discriminate between several single-field
inflationary models using this new, high quality, dataset
\cite{alabidi,PeirisEasther,Lewis:2006ma,Seljak:2006bg,Magueijo:2006we,
Liddle06,kinney06,Martin:2006rs}.
One of the main conclusions of these works is that some inflationary
models, such as quartic chaotic models of the
form $V(\phi)\sim \lambda \phi^4$, may be considered ruled out by the
current data, while others, such as chaotic inflation with a quadratic
potential $V(\phi) \sim m^2 \phi^2$, are consistent with all data sets. 

One important assumption in these analyses (apart for 
\cite{Martin:2006rs}) is that the
inflaton's potential is featureless, i.e., there is no
preferred scale during inflation and the primordial power
spectrum of density perturbations in Fourier $k$-space can be well 
approximated by a power law $k^n$, where the spectral index $n$ is
almost scale independent.
The main prediction of these models is that the anisotropy angular
power spectrum should be ``smooth'' and not show features in addition
to those provided by the baryon-photon
plasma oscillations at decoupling within the framework of the
standard $\Lambda$CDM model of structure formation.

The current WMAP data is in very good agreement with this
hypothesis: several non-standard features in the anisotropy angular
power spectrum detected in the first year data have now disappeared
thanks to the longer integration time of the observations and better
control of systematics (see \cite{wmap3cosm}). 

However, features in the large scale anisotropy spectrum 
are still present in the new release. Moreover, some of the
cosmological parameters derived from the new WMAP data, like, for
instance, the low value of the variance of  fluctuations $\sigma_8$,
appear in tension with those derived by complementary data sets. It is
therefore timely to investigate a larger set of inflationary models
and to consider a cosmological origin of these unexpected features.

A departure from power law behavior of the primordial power spectrum 
could be caused by a change of the initial conditions, due
to trans-planckian physics \cite{Martin:2003kp,Easther:2002xe}
or unusual initial field dynamics \cite{Burgess:2002ub, Contaldi:2003zv} 
or by some brief violation of the slow roll conditions during 
inflation~\cite{Kofman:1989ed,Starobinsky:1992ts}.
We will investigate a model of the second type where 
features in the temperature and density power spectra
arise due to a step-like change in the potential parameters, 
as proposed by~\cite{ace}. A sharp step in the inflaton mass, 
caused, e.g., by a symmetry breaking phase transition, generates 
indeed $k$-dependent 
oscillations in the spectrum of primordial density perturbations.

The goal of our paper is to make use of the recent 
three year WMAP data (WMAP3) and other datasets to constrain the 
possibility of a step feature in the inflaton potential. 
For this purpose we adopt the phenomenological model proposed by 
Adams et al. \cite{ace}, where a step feature is added to the 
chaotic inflationary potential in the following way:
\begin{equation}
V(\phi) = {1\over 2} m^2 \phi^2 \,
\left( 1 + c \tanh \left( \frac{\phi-b}{d} \right) \right),
\label{tan-potential}
\end{equation}
where $m$ is an overall normalization factor, $c$ determines the
height of the step, $d$ its gradient and $b$ is the field value on
which the step is centered.
Previous phenomenological studies of the same~\cite{Peiris:2003ff} 
or other oscillatory features~\cite{Martin:2004yi,Kawasaki:2004pi} 
have been limited to the first year WMAP data, and in general 
a full analysis varying also all cosmological parameters is 
still missing and is the major result of this work.

The paper is organized as follows: In Sec.\ II we briefly review step-inflation models. In Sec.\ III we describe our analysis method. 
In Sec.\ IV we present our results and, finally, in Sec.\ V we
derive our conclusions.

\section{Inflation models with a step in the potential.}

Inflationary models with a step can naturally arise in theories with 
many interacting scalar fields, e.g., in supergravity models. 
In general, these models contain several flat directions in 
field space and thus offer the possibility 
to have multiple inflationary phases separated by phase transitions 
\cite{ars97}, or even inflation with a {\it curved} trajectory in field 
space \cite{mukh98,star01,groot01}. 
In the last case the presence of additional {\it active} scalar degrees of 
freedom generates not only the adiabatic mode of curvature perturbations,
but also the isocurvature one.
Since the data do not seem to require an isocurvature
component~\cite{isobound}, we will restrict ourselves to the case where
the phase transition does 
not appreciably change the rolling direction corresponding to the 
inflaton field and the energy density is always dominated by a single
field.
Also, we will investigate the simplest scenario and assume that the 
sole effect of the phase transition is to change the parameters of 
the Lagrangian for the inflaton field, in particular its mass.

Consider a hybrid inflationary potential of the type
\begin{equation}
V = V_0 + {1\over 2} m_0^2 \phi^2 + {\lambda^2 \over 4} \left( \psi^2 - M^2 \right)^2 
+ \lambda^2 \phi^2\psi^2\; ,
\end{equation}
where $\phi$ is the inflaton field, while $\psi $ is a hybrid field
that takes a vacuum expectation value during inflation when the
inflaton reaches the critical value $\phi_c^2 = { M^2 \over 2}$.
In the usual case of hybrid inflation, this transition is so
strong that it stops the inflationary phase. But, if the coupling
$\lambda $ is sufficiently small, the back-reaction of $\psi $ is too
weak and inflation continues.
On the other hand, the parameters of the Lagrangian change their value
and for example the inflaton effective mass becomes
\begin{equation}
m_{\text{eff}}^2 (\phi) = 
m_0^2 + \lambda^2 \langle \psi^2 \rangle (\phi )\; .
\end{equation}
In this scenario, even if the classical inflaton is nearly
unperturbed, the inflaton perturbations are affected and the
primordial power spectrum is modified.\\

It is well known that a step in the mass generates oscillations in the 
primordial spectrum and this can be described analytically in the WKB 
approximation \cite{hs04}.
The behavior of the inflaton mass is determined by the dynamics of the
phase transition and the growth of the hybrid field fluctuations, 
which become tachyonic after the critical point. This growth 
depends on the classical inflaton field motion, but is 
in general so fast that $\psi $ reaches the minimum in a very 
small number of $e$-foldings \cite{fggklt01,abc01-hyb,cpr02,ggg02}. 
The inflaton mass is therefore reasonably well approximated by a 
hyperbolic tangent, and so we take
\begin{equation}
m_{\text{eff}}^2 (\phi) \simeq m^2 \left( 1 + c \tanh \left( \frac{\phi-b}{d} \right) \right).
\end{equation}
Here, we see that the parameter $m$ is an average inflaton mass and 
$b$ is of the order of the critical value $\phi_c$.
The other two parameters determine the duration of the transition and
the strength of the effect on the inflaton's mass.
Note that we work in Planck units, so all dimensional
quantities like $b$ and $d$ should be multiplied by $M_{\text{P}}$ in order to
obtain their value in physical units. 
In this article we restrict ourselves to the case of chaotic 
inflation, where the mass term determines both the classical dynamics
of the inflaton and the behavior of the perturbations. A more general
discussion will be left for a longer publication \cite{chmss2}.
We therefore assume that $V_0$ and also its change 
due to the phase transition are completely negligible. 
We also discuss here only positive $c$ values;
negative $c$ is also allowed, but is restricted to be very small
to avoid the presence of another minimum in the potential away
from $\phi =0$.

Since we cannot rely on the slow roll approximation for a generic
choice of parameters, we integrate the equations for the background
and for the modes numerically as discussed in detail in~\cite{ace}.
The equations for the inflaton field and the Hubble parameter in
Planck units are simply
\begin{eqnarray}
\ddot \phi + 3 H \dot\phi + V'(\phi) = 0 \label{eom1}\\  
3 H^2 = {\dot\phi^2\over 2} + V(\phi).\label{eom2}
\end{eqnarray}
We assume slow roll as the initial condition for $\phi \gg \phi_c$ and solve the evolution numerically until the end of inflation in
order to determine the number of $e$-foldings between $b$ and the
end of inflation. 

The equation for the Fourier components of $u = - z {\cal R}$, the curvature 
perturbation \cite{Stewart:1993bc}, takes the usual form
\begin{equation}
u_k'' + \left( k^2 -{z'' \over z} \right) u_k = 0 \label{eom3}
\end{equation}
where $z = a \dot\phi /H$ is given by the background dynamics and the
primes and dots denote derivatives with respect to conformal time and
physical time, respectively.
Using the equations of motion for the classical field, we have in general
\begin{equation}
{z''\over z} = 2 a^2 H^2 \left(1 +  {7\dot\phi^2 \over 4 H^2} +
{\dot\phi^4 \over 4 H^4} + {V' \dot\phi\over H^3}- {V'' \over 2 H^2} 
\right),
\end{equation}
where $V'$ and $V''$ are the derivatives of the potential with respect to the
inflaton field. In the step model, this quantity can deviate
substantially from the slow-roll expectation \mbox{$z''/z \simeq 2 a^2
H^2$} at the time of the transition. We solve equations
\mbox{(\ref{eom1}) to (\ref{eom3})} numerically using a Bulirsch-Stoer
algorithm for free field initial conditions at an initial time when
\mbox{$k^2 \gg  z''/z$} is satisfied.

Once we know the solution for the mode $k$, we can determine the
primordial power spectrum
\begin{equation}
\mathcal{P_R} (k) =  {k^3\over 2\pi} \left| \frac{u_k}{z} \right|^2,
\end{equation}
evaluated when the mode is well outside the horizon. 
Our results are stable with respect to changes in the exact time when
we set the initial conditions and when we compute the spectrum.

\begin{figure}
\includegraphics[width=2in,angle=-90]{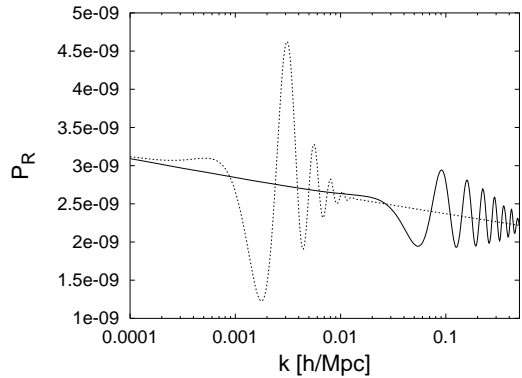}
\caption{\label{jan1} Effects of a step in the potential
on the power spectrum of curvature perturbations.
Here we show the primordial spectrum for the two best
fit points, corresponding to \mbox{$b=14.81$}, \mbox{$c=0.0018$}, \mbox{$d=0.022$} (dashed line)  
and \mbox{$b=14.34$}, \mbox{$c=0.00039$}, \mbox{$d=0.006$} (solid line).}
\end{figure}

The resulting spectra as a function of $k$ for different 
parameters are shown in Figure~\ref{jan1}. Essentially, the spectrum
shows a power-law behavior with a superimposed oscillation.

How will the four parameters of our model affect the shape of the
spectrum?
The overall normalization of $\mathcal{P_R}$ is proportional to $m^2$,
$b$ determines the wavelength at which the feature appears and the
maximum amplitude of the oscillations is roughly proportional to $c$.
Generally, the dominant contribution to $z''/z$ comes from the $V''$
term and is proportional to $c/d^2$, so the range of $k$ affected by
the feature depends on the square root of $c/d^2$.
Note that away from the step, the slow roll conditions are satisfied 
and the spectrum recovers the usual power law form with spectral index
given by $ n_s = 1-2/N \simeq 0.96$ for a number of $e$-foldings $N$
equal to 50, which is what we assume for our analysis. 
The value of the spectral index is the same before and after the step
since it does not depend on $m^2$ in $m^2 \phi^2$ models.
Also, for values of the parameters where the slow-roll conditions 
are always satisfied (i.e., small values of $c/d^2$), the spectrum does
not show a full oscillation, but a dip at the scales corresponding to
the transition. So even in this case it is not so well approximated by
the usual power-law with a constant spectral index.

\section{CMB Analysis}
\label{secCMBanalysis}

We compare the theoretical model described in the previous sections
with a set of current cosmological data by making use of 
a modified version of the publicly available Markov Chain 
Monte Carlo (MCMC) package \texttt{cosmomc}~\cite{Lewis:2002ah}.

We sample an eight-dimensional set of parameters. Four of them
determine the primordial power spectrum, namely the $b,c$ and $d$
parameters of the step-inflation model as described in the previous
section and the overall normalization of the primordial power spectrum
$A_S$ (equivalent to $m^2$ as discussed earlier). The remaining four
cosmological parameters are the physical baryon and CDM densities, 
$\omega_b=\Omega_bh^2$ and $\omega_c=\Omega_ch^2$, the ratio of 
the sound horizon to the angular diameter distance at decoupling,
$\theta_s$ and finally, the optical depth to reionization, $\tau$.
Furthermore, we consider purely adiabatic initial conditions, impose
flatness and neglect neutrino masses. 

We include the three year data \cite{wmap3cosm} (temperature and 
polarization) using the likelihood routine for supplied 
by the WMAP team and available at the \texttt{LAMBDA} web
site.\footnote{http://lambda.gsfc.nasa.gov/} 
We marginalize over the amplitude of the Sunyaev-Zel'dovich signal.
The MCMC convergence diagnostics are done on four chains using the
Gelman and Rubin ``variance of chain means''$/$``mean of chain variances''
$R$ statistics for each parameter, demanding that $R-1 < 0.1$.  Our
$2D$ constraints are obtained after marginalization over the
remaining ``nuisance'' parameters, again using the programs included
in the \texttt{cosmomc} package.

In addition to CMB data, we also consider the constraints on the
real-space power spectrum of galaxies from the Sloan Digital Sky
Survey (SDSS) \cite{thx}.

We restrict the analysis to a range of scales over which the 
fluctuations are assumed to be in the linear regime 
($k < 0.2 \; h/{\text{Mpc}}$). When combining the matter power spectrum 
with CMB data, we marginalize over an additional nuisance parameter
$b'$, the dark versus luminous matter bias. Furthermore, we make use of the 
HST key project measurement of the Hubble parameter $H_0 = 100h \text{
km s}^{-1} \text{Mpc}^{-1}$ \cite{hst} by multiplying the likelihood by a
Gaussian centered around $h=0.72$ with a standard deviation $\sigma =
0.08$. Finally, we impose a top-hat prior on the age of the universe,
$10 < t_0 < 20$ Gyrs, and a Gaussian prior on $\Omega_b h^2$ centered
around $0.022$ with a standard deviation of $0.002$ from BBN
constraints, cf.~Ref.~\cite{Lewis:2002ah}.

We demand that the feature appear at a wavelength to which our data is
sensitive, so our analysis will be limited to the interval
\mbox{$13.5 < b < 16$}. Apart from that we also impose logarithmic
priors on the other step parameters: \mbox{$\log c \in[-6,-1]$},
\mbox{$\log d \in [-2.5,-0.5]$} and \mbox{$\log c/d^2 \in [-5,3]$} and
flat priors on the cosmological parameters.

As it turns out, the likelihood distribution $\mathcal L$ has a rather
odd shape and some of the interesting features are at low
likelihoods. In order to improve mixing and get a better coverage of
the low likelihood regions, we sample $\mathcal{L}^{1/3}$ instead of
$\mathcal L$ (i.e., we use ``heated'' chains at $T=3$).

As a measure of the performance of the step model, we compare its best
fit $\chi^2$ with the best fit $\chi^2$ of a reference model, which we
take to be the ``vanilla'' 6 parameter ($\Omega_b h^2$, $\Omega_c h^2$,
$\theta_s$, $\tau$, $A_S$ and $n_s$) power law $\Lambda$CDM model.

\section{Results}

\begin{figure}
\includegraphics[width=.5\textwidth]{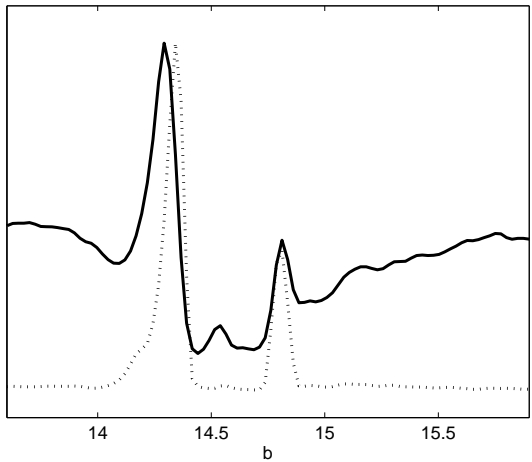}
\caption{\label{likelihoodwmaponly} 
Marginalised likelihood (solid line) and projection of the likelihood
distribution (dotted line) for the $b$ parameter in the case of WMAP
only. Two peaks for $b$ at $b=14.3$ and $b=14.8$ are clearly
visible. The one at $b=14.8$ provides a good fit to the low $\ell$
WMAP glitches. It is evident that the likelihood function is
far from gaussian in this direction. The difference between the two
curves is caused by a volume effect when integrating over the other
parameter directions.} 
\end{figure}

\begin{figure}
\includegraphics[height=.5\textwidth,angle=-90]{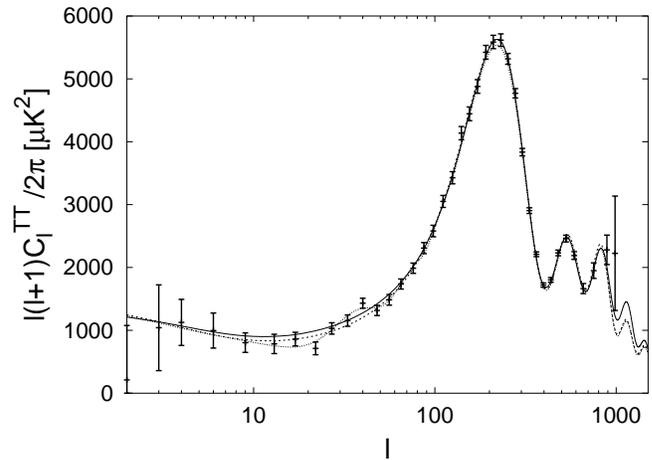}
\caption{\label{wmapcltt} 
This plot shows the temperature anisotropy angular power spectrum
of the best fit step model (WMAP only, solid line) and, for reference,
the best fit 6 parameter power law $\Lambda$CDM model (dashed line).
The dotted line shows the effect of a feature near $b=14.8$ for WMAP
data only, i.e., the ``local'' best fit at the lower peak in Figure 
\ref{likelihoodwmaponly}.}
\end{figure}

Let us first consider the WMAP dataset alone.
In Figure \ref{likelihoodwmaponly} we plot the mean 
likelihood for the $b$ parameter which determines the position (scale)
of the step in the potential. If some value of $b$ is
preferred by the data, then it would hint at the presence of a feature.
As we can see, the mean likelihood distribution clearly indicates two
maxima for the $b$ parameter at $b=14.3$ and $b=14.8$,
respectively. The feature at $b=14.8$ is able to produce a good fit to
the WMAP low-$\ell$ glitches (yielding an improvement of $\Delta
\chi^2 \simeq 5$ over the vanilla model) and agrees with the results
of Ref.~\cite{Peiris:2003ff} for fixed cosmological parameters and
WMAP first year data. Interestingly, the authors of
\cite{Shafieloo:2003gf} find an oscillating feature at roughly the  
same scale by reconstructing the primordial power spectrum from the 
first year data. The minimum at $b=14.3$ provides oscillations on
smaller scales ($\Delta \chi^2 \simeq 7$), beyond the second peak in
the anisotropy power spectrum, see Figure~\ref{wmapcltt}.

\begin{figure}[h!]
\includegraphics[width=0.5\textwidth]{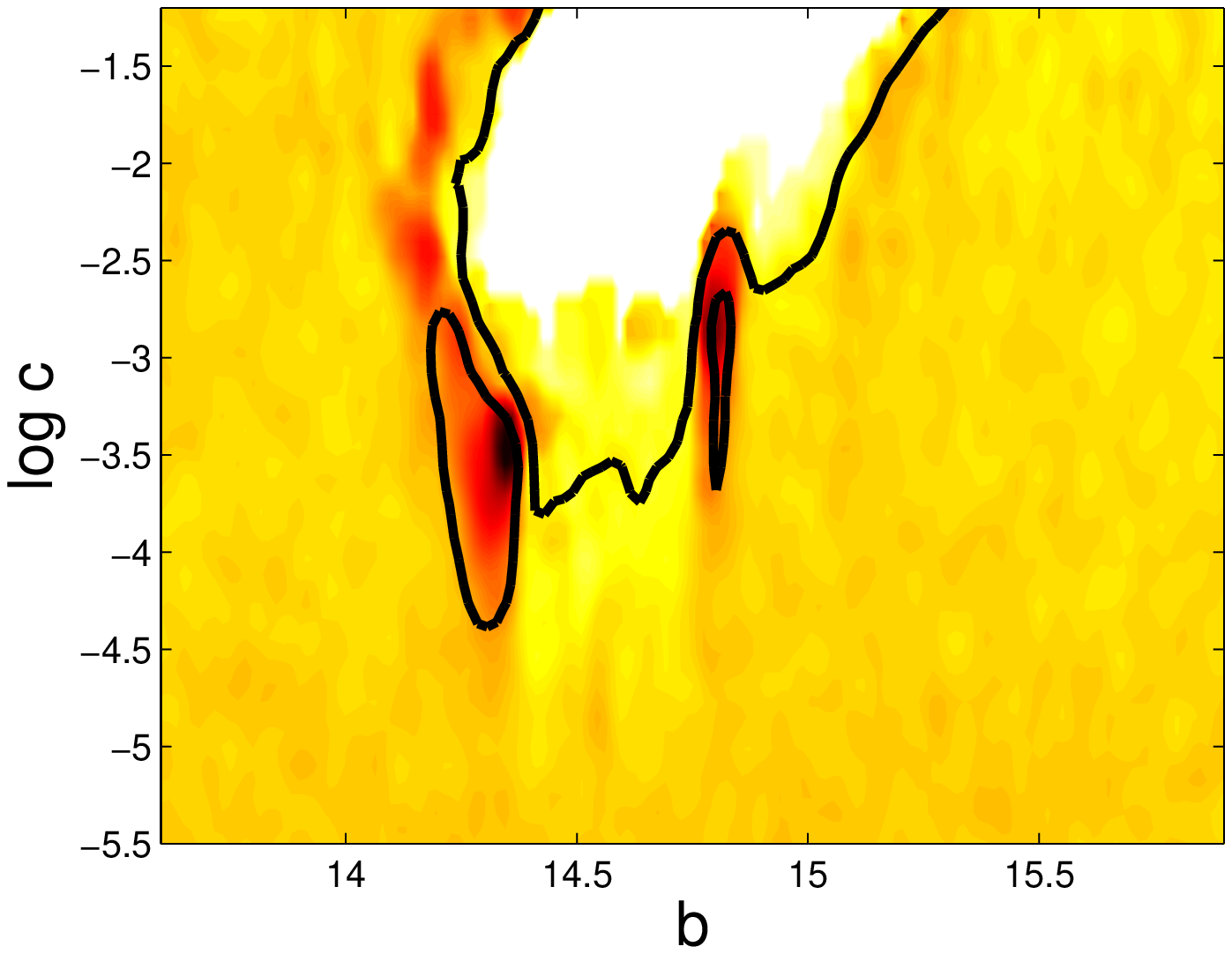}
\includegraphics[width=0.5\textwidth]{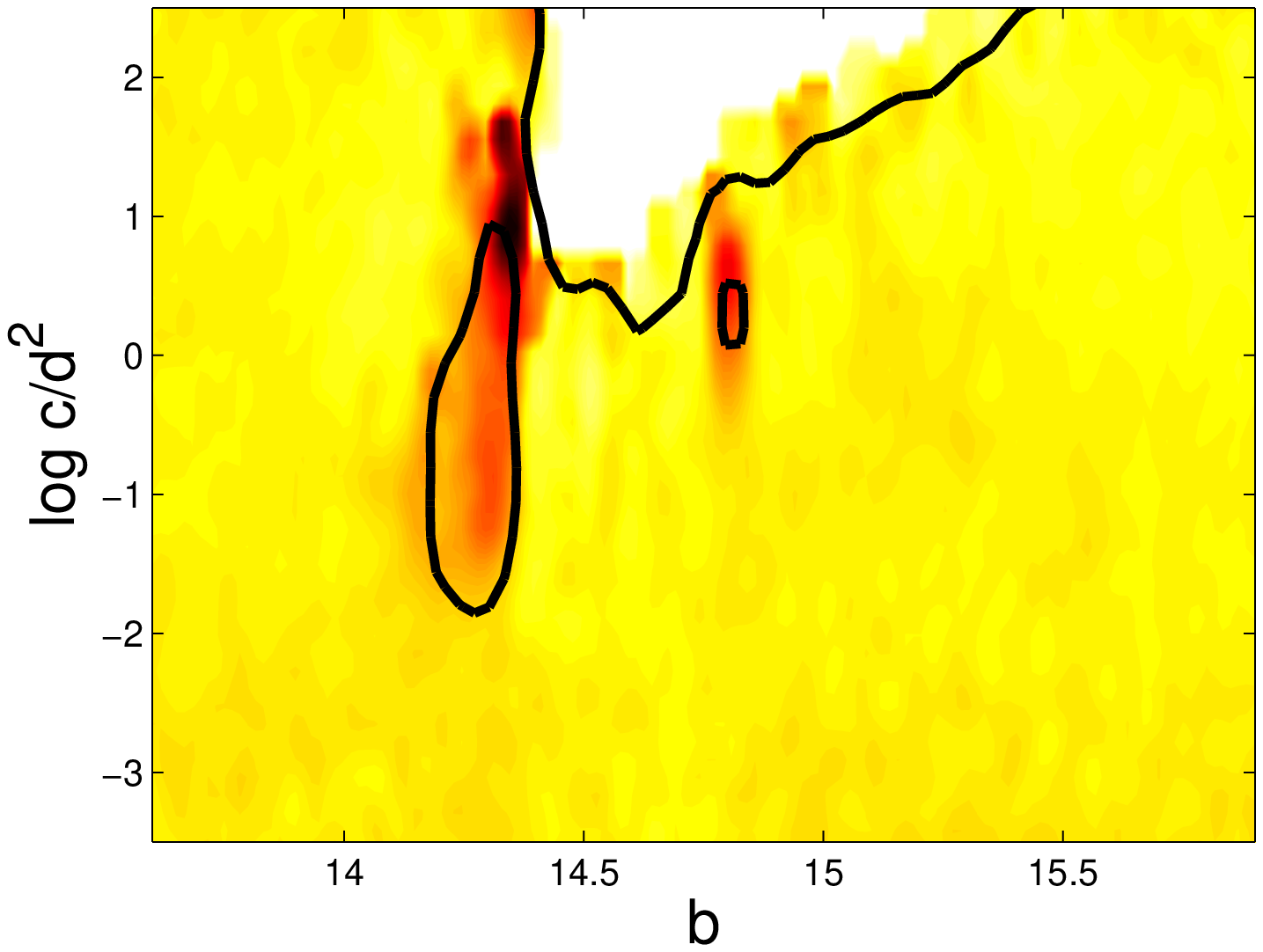}
\caption{\label{contour} 
Mean likelihood and marginalized likelihood contours in the ($b,\,c$)
and $(b,\,c/d^2)$ planes at $8 \%$ and $99 \%$ c.l.~for WMAP data
only. The peaks comprise less than ten per cent of the total volume of
the likelihood function.}
\end{figure}

It is interesting to project the likelihood function onto the
$(b,\,\log c)$ plane (Figure \ref{contour}, upper panel).
First of all, we see that for a range of values ($14.3 < b <15.5$) a
region of step models with $\log c$ $< -3$ is ruled out at $99 \%$
confidence level. This disfavored region corresponds to the region in
$k$ space that is better sampled by the WMAP data, and where,
therefore, the data provide the strongest constraints. Secondly, the
two aforementioned maxima in $b$-space can again be seen in the $2D$
projection. These maxima are close to the two boundaries (large and
small scales) of the region sampled by WMAP and centered around
amplitude $\log c \sim - 3$.
We find that the WMAP polarization and cross temperature-polarization
data are rather insensitive to the presence of features.

The $d$ parameter is not well constrained by the data due
to a degeneracy with $c$. Instead, we consider the constraints in the
($b,\,\log c/d^2$) parameter space (Figure \ref{contour}, bottom
panel). This parameter is as well constrained as $c$ and, again, the
presence of two maxima for $b$ is evident. Also, the maxima are at
values of $c/d^2$ of order $1$, where the slow-roll conditions are
strongly violated; values of $c/d^2 \ll 0.1$ correspond, on the other
hand, to the usual slow-roll $m^2 \phi^2$ inflation and cannot be
excluded by the data.

Note that in the ($b,c,d$)-subspace of parameter space the likelihood
function is very oddly shaped and can by no means be approximated by a
multivariate Gaussian. As a consequence, the likelihood at the
boundaries of this subspace is generally not negligible (see, e.g.,
Figure \ref{likelihoodwmaponly}). In fact, the likelihood function
will have a large plateau of constant likelihood in regions where the
step model cannot be distinguished from the $m^2 \phi^2$ chaotic inflation
model, either because the step is too small or too smooth (small $c$,
large $d$) or because the feature appears at wavelengths the data is
not sensitive to ($b$ too small or too large). Apart from the two
peaks, we also find a valley that can be excluded at a high confidence
level. However, since the plateau may contain a significant fraction of the
total volume of the likelihood distribution, constraints derived from
marginalization will be dependent on the priors on $b$, $c$ and $d$.

\begin{figure}[t!]
\includegraphics[width=.5\textwidth]{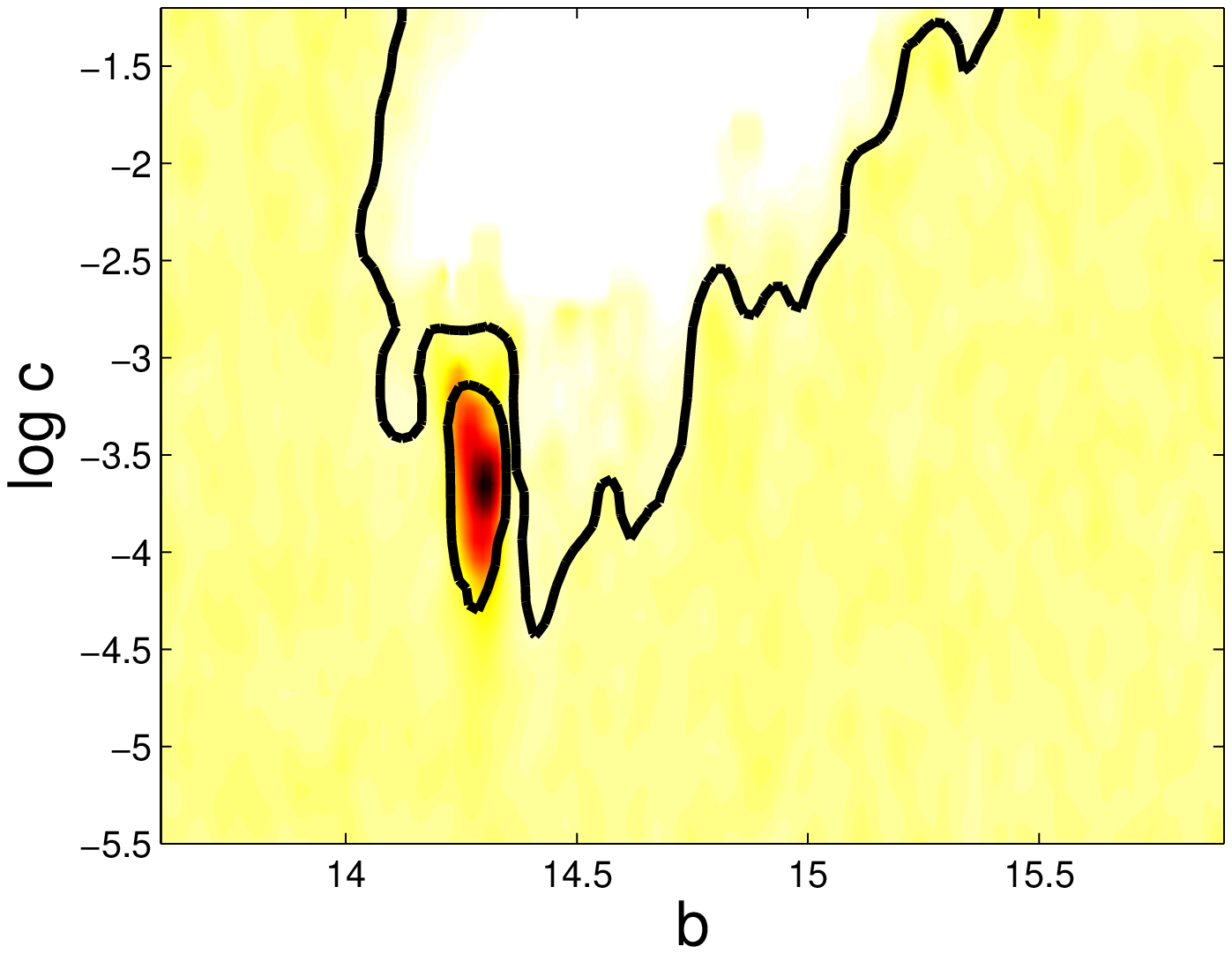}
\includegraphics[width=.5\textwidth,height=185pt]{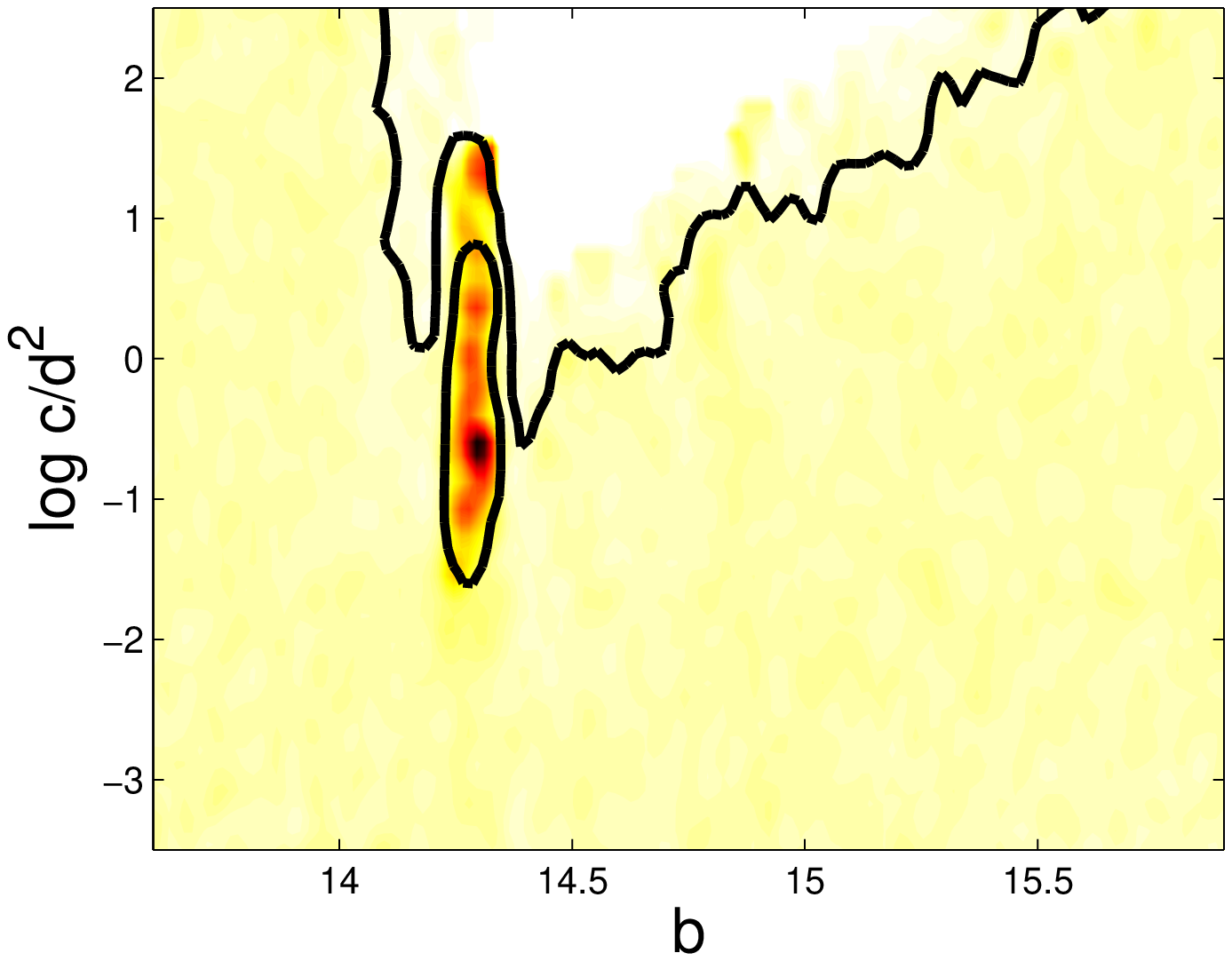}
\caption{\label{contour2} 
Likelihood contours in the $(b,\,c)$ and $(b,\,c/d^2)$ planes 
at $68 \%$ and $99 \%$ confidence level for CMB+SDSS.}
\end{figure}

For our choice of priors and WMAP data only, we find that the peak
regions contribute about $8 \%$ of the total volume, so, from a
Bayesian standpoint, the WMAP data alone do not require the presence
of a feature.


Figure \ref{wmapcltt} indicates that the best fit model has a feature at a
range of wavelengths where the WMAP data are limited by large
systematic errors. It is therefore interesting to enquire whether the
inclusion of other data sets which are more sensitive at small scales
will corroborate this result.
To this end we add small scale CMB data from the ACBAR, BOOMERANG,
CBI, MAXIMA and VSA experiments \cite{acbar,boom,cbi,maxi,vsa} and
the SDSS large scale structure data.

These data sets probe mainly smaller scales and are not sensitive to the
large scale feature at $b=14.8$.

\begin{figure}[t!]
\includegraphics[height=0.5\textwidth,angle=270]{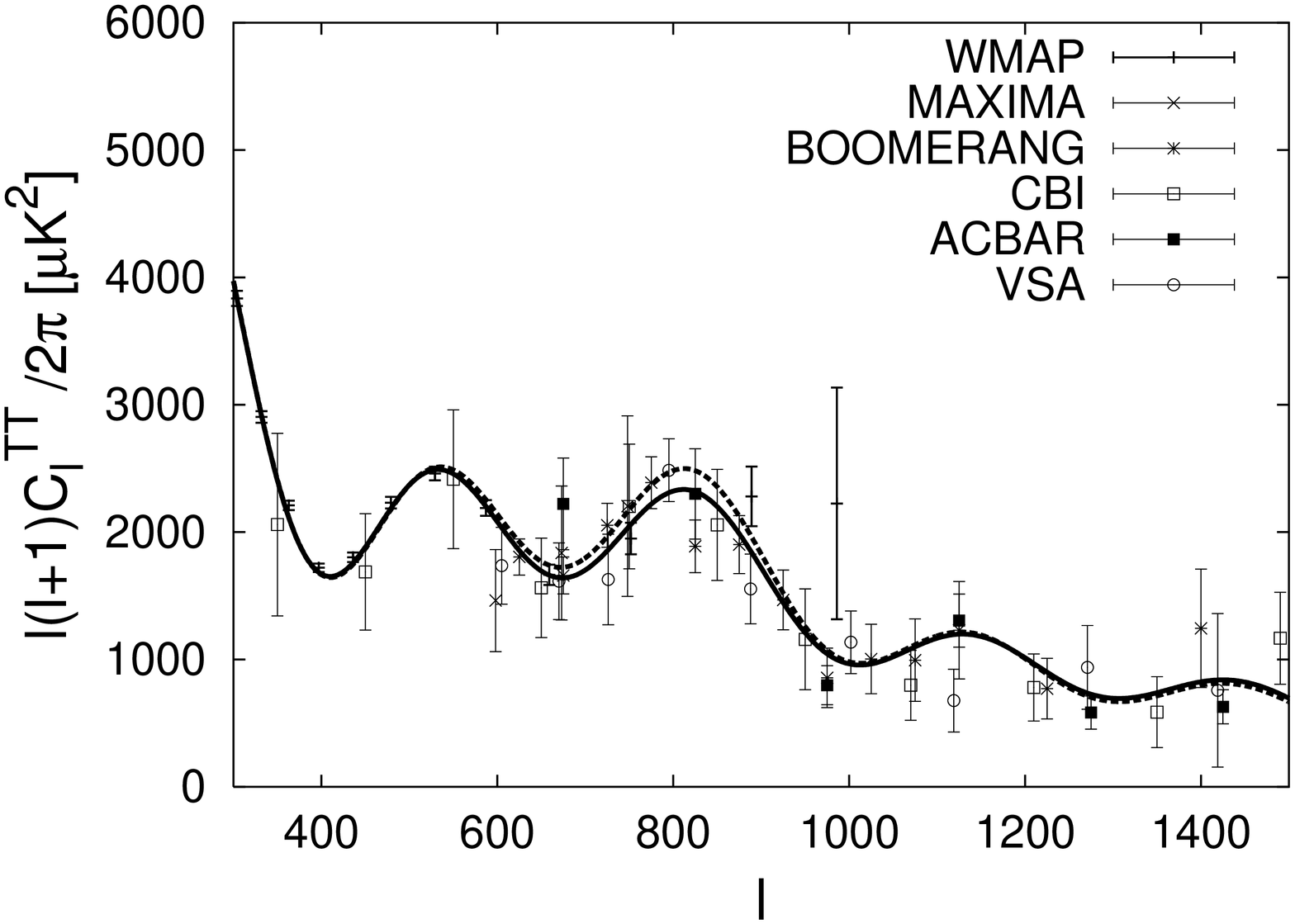}
\includegraphics[height=0.5\textwidth,angle=270]{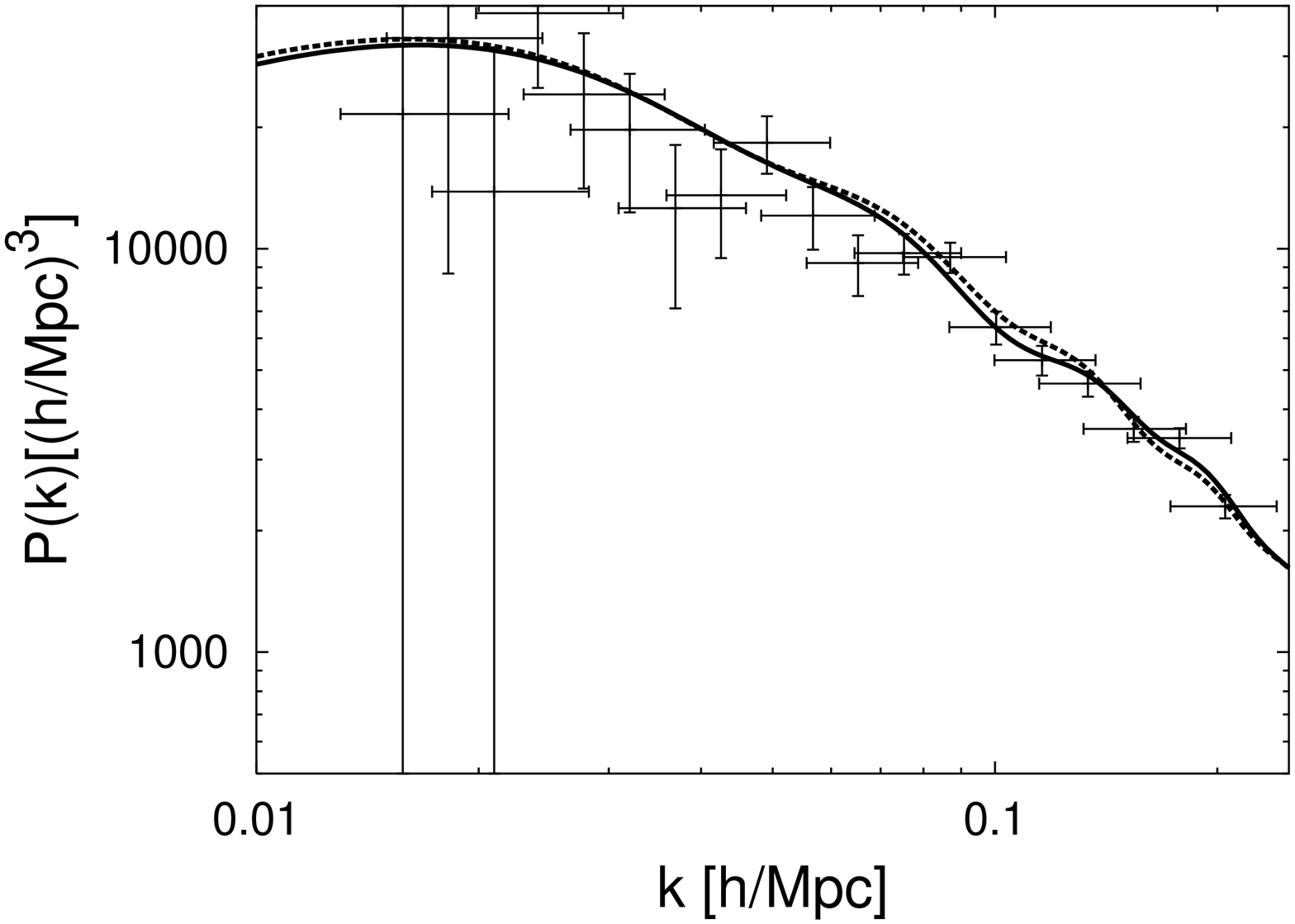}
\caption{\label{wmap+all} 
Top: temperature anisotropy angular power spectrum with small scale
CMB data.\\
Bottom: galaxy power spectrum and the SDSS data.\\
The solid lines depict the best fit step model, the dashed
line the best fit reference model.}
\end{figure}

We find that including these small scale data improves the constraints on the
oscillations  quite significantly: $\Delta \chi^2
\simeq 15$, to which the SDSS data alone contribute about 6.
As we can see from Figure \ref{wmap+all} (bottom panel), 
step-induced oscillations in the matter power spectrum seem to provide
a much better fit to the SDSS data.

In Figure \ref{contour2} we plot the likelihood contours in the
($b,\log c$) plane at $68 \%$ and $99 \%$ c.l.~for the CMB+SDSS
case. Adding the SDSS data increases the statistical significance of
the maximum, the peak near ($b \simeq 14.3$ , $\log c \simeq -3.5$,
$\log c/d^2 \simeq 0$) now contains about $70\%$ of the total
volume of the likelihood function, the likelihood at the boundaries of
parameter space is suppressed and hence, the results are much less prior
dependent. Also, we can rule out a much larger chunk of parameter
space for $b<14.5$ due to the increased sensitivity of the data on the
corresponding scales.

The variance test convergence stats using last half chains
are $0.0007$ for $b$, $0.0038$ for $\log c$ and $0.0036$ for  
$\log c/d^2$, showing a robust convergence of the 
chains.

We also considered data from the 2dF Galaxy Redshift Survey 
\cite{Cole:2005sx} and we found that the CMB+SDSS results are
stable under the inclusion of the 2dF dataset.

Of course, oscillatory behavior in the observed data could well have
its origin in uncorrected or unidentified systematic effects such as
a scale dependent bias. However, as indicated by our results, the
presence of multiple step-like features in the inflaton potential
(expected from, e.g., supergravity or M-theory models
\cite{Ashoorioon:2006wc}) is also a viable solution.

It is important to check if the step in the potential suggested by
the CMB+SDSS analysis has some impact on the estimation of the
remaining cosmological parameters. We find no correlations
between $b, c, d$ and the cosmological parameters 
$A_S, \theta_s$ and $\tau$. This is clearly due to
the fixed spectral index the model has away from the feature.
 
\begin{figure}[h!]
\includegraphics[width=.5\textwidth]{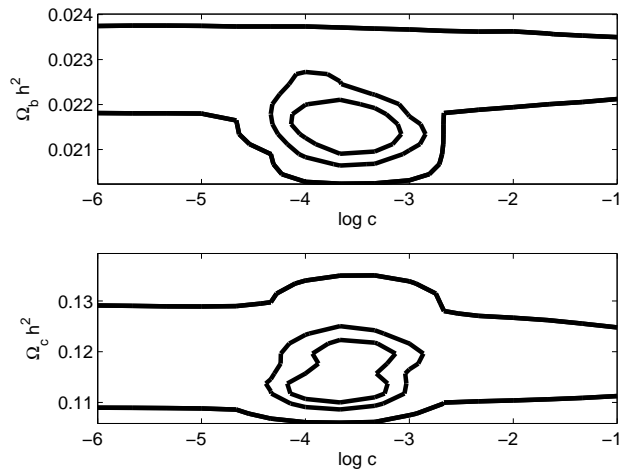}
\caption{\label{ccorr} 
Correlations between $c$ and the cosmological parameters
$\Omega_b h^2$ and $\Omega_c h^2$ for CMB+SDSS. 
Plotted are the $50 \%$, $68 \%$ and $99 \%$ confidence levels. The
best fit step model peak prefers a slighly lower value of $\Omega_b
h^2$ than the plateau region where the feature is insignificant. One
can also see that near the best fit region, the larger $c$ the
smaller the corresponding value of $\Omega_b h^2$.}
\end{figure}

On the other hand some correlation is present between $c$
and the baryon and cold dark matter energy densities $\Omega_bh^2$ 
and $\Omega_ch^2$. Figure \ref{ccorr} shows the confidence level
likelihood contours in the $(\Omega_b h^2,c)$ and $(\Omega_c h^2,c)$
parameter space. 
A deeper step in the potential (larger $c$) has the effect of
making the data more compatible with a lower baryon density and
a higher cold dark matter density. While the effect is small,
it is interesting to note that the baryon density derived from the
WMAP data in the framework of the standard model is generally 
larger than that predicted by standard big bang nucleosynthesis and
measurements of the primordial deuterium abundance (see,
e.g., \cite{fields}).

\pagebreak

\section{Mimicking Baryonic Oscillations}

Recently, a detection of oscillations in the correlation function of
the Luminous Red Galaxies sample of the SDSS has been reported in
\cite{eisenstein}. If those oscillations are the imprint of primordial
acoustic oscillations in the primeval baryon+photon plasma, then they
may provide a standard ruler at the redshift of the survey and a very
powerful tool for testing the late time evolution of the universe and,
ultimately, the appearance of dark energy. Clearly, if oscillations in
the primordial spectrum are present they may mimic baryonic
oscillations \cite{Wang:2002hf} at different scales and drastically
change the estimation of cosmological parameters.  

\begin{figure}[h!]
\includegraphics[width=.5\textwidth]{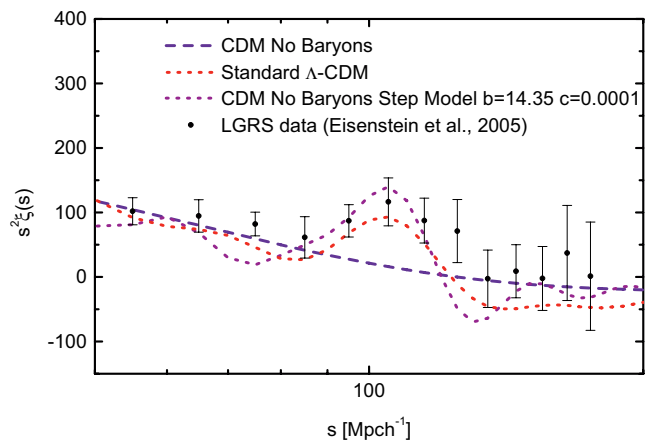}
\caption{\label{bao} 
Correlation functions for a model with no baryons,
the standard $\Lambda$CDM model ($\Omega_b=0.05$) and
a model with no baryons but with a step in the inflationary potential.
The data points are from the LRGS sample of \cite{eisenstein}.}
\end{figure}

As a qualitative example, we plot in Figure \ref{bao} the correlation functions
for a model with no baryons, the standard $\Lambda$CDM model
($\Omega_b=0.05$) and a model with no baryons but with a step in the
inflationary potential. As we can see, a model with no baryons but
with oscillations in the primordial spectra can reproduce the observed
data very well. A more detailed and quantitative 
analysis will be presented in a future paper \cite{chmss2}.
  
\section{Conclusions}

The new three year WMAP data seem to confirm the presence of
non-standard large
scale features on the cosmic microwave anisotropy power spectrum.
While these features may hint at uncorrected experimental
systematics, a possible cosmological way to generate large angular 
scale oscillations is to introduce a sharp step in the inflaton 
potential.
By making use of current cosmological data we derive constraints on 
the position, magnitude and gradient of a possible step in the
inflaton potential. Our conclusion is that such a step, 
while strongly constrained by current data, is still allowed and may
provide an interesting explanation to the current measured deviations
from the standard featureless spectrum at low $\ell$.

Surprisingly though, the combination of all CMB data sets with the
SDSS data seems to prefer a feature at small scales, which could mimic
the effect of baryonic oscillations and reduces the best fit value of
$\Omega_b$. Note that for this it is sufficient to have a minute
change, of order $0.1\%$, in the inflaton mass parameter, but a 
relatively fast one.

It is an open question if such a sharp step can be realized in
realistic inflationary models and if the effect is physical. 
In general we can exclude the presence of strong features with $c \geq
0.003 $ in the observable range. 

Future experiments like PLANCK will provide better measurements
of the polarization and cross temperature-polarization spectra,
providing an important check for possible non-standard features.

\acknowledgments
We would like to thank S.~Sarkar for useful discussions. JH thanks
Y.~Wong for discussions and comments on the manuscript. LC and JH
acknowledge the support of the ``Impuls- und Vernetzungsfonds'' of the
Helmholtz Association, contract number VH-NG-006.




\begin{thebibliography}{99}
\frenchspacing

\bibitem{wmap3cosm} 
D.N.~Spergel {\it et al.},
arXiv:astro-ph/0603449.

\bibitem{wmap3pol} 
L.~Page {\it et al.},
arXiv:astro-ph/0603450.

\bibitem{wmap3temp}
G.~Hinshaw {\it et al.},
arXiv:astro-ph/0603451.

\bibitem{wmap3beam} 
N.~Jarosik {\it et al.},
arXiv:astro-ph/0603452.


\bibitem{Starobinsky:1979ty}
A.A.~Starobinsky,
JETP Lett.\  {\bf 30} (1979) 682
[Pisma Zh.\ Eksp.\ Teor.\ Fiz.\  {\bf 30} (1979) 719].

\bibitem{muk81} 
V.F.~Mukhanov and G. V. Chibisov, 
JETP  Lett. {\bf 33} (1981) 532.

\bibitem{bardeen83} 
J.M.~Bardeen, P. J. Steinhardt, and M. S. Turner, 
Phys. Rev. D {\bf 28} (1983) 679.

\bibitem{alabidi} 
L.~Alabidi and D.H.~Lyth,
arXiv:astro-ph/0603539.

\bibitem{PeirisEasther}
H.~Peiris and R.~Easther,
arXiv:astro-ph/0603587;
R.~Easther and H.~Peiris,
arXiv:astro-ph/0604214.

\bibitem{Lewis:2006ma}
A.~Lewis, 
arXiv:astro-ph/0603753.

\bibitem{Seljak:2006bg}
U.~Seljak, A.~Slosar and P.~McDonald,
arXiv:astro-ph/0604335.

\bibitem{Magueijo:2006we}
J.~Magueijo and R.D.~Sorkin,
arXiv:astro-ph/0604410.

\bibitem{Liddle06}
D.~Parkinson, P.~Mukherjee and A.R.~Liddle,
arXiv:astro-ph/0605003;
C.~Pahud, A.R.~Liddle, P.~Mukherjee and D.~Parkinson, 
arXiv:astro-ph/0605004.

\bibitem{kinney06}
W.H.~Kinney, E.W.~Kolb, A.~Melchiorri and A.~Riotto,
arXiv:astro-ph/0605338; PRD in press.  

\bibitem{Martin:2006rs}
J.~Martin and C.~Ringeval,
arXiv:astro-ph/0605367.

\bibitem{Martin:2003kp}
R.H.~Brandenberger and J.~Martin,
Mod.\ Phys.\ Lett.\ A {\bf 16} (2001) 999
[arXiv:astro-ph/0005432];
J.~Martin and R.~Brandenberger,
Phys.\ Rev.\ D {\bf 68} (2003) 063513 
[arXiv:hep-th/0305161].

\bibitem{Easther:2002xe}
R.~Easther, B.R.~Greene, W.H.~Kinney and G.~Shiu,
Phys.\ Rev.\ D {\bf 66} (2002) 023518
[arXiv:hep-th/0204129].

\bibitem{Burgess:2002ub}
C.P.~Burgess, J.M.~Cline, F.~Lemieux and R.~Holman,
JHEP {\bf 0302} (2003) 048
[arXiv:hep-th/0210233].

\bibitem{Contaldi:2003zv}
C.R.~Contaldi, M.~Peloso, L.~Kofman and A.~Linde,
JCAP {\bf 0307} (2003) 002
[arXiv:astro-ph/0303636].

\bibitem{Kofman:1989ed}
L.~Kofman, G.R.~Blumenthal, H.~Hodges and J.R.~Primack,
ASP Conf.\ Ser.\  {\bf 15} (1991) 339.


\bibitem{Starobinsky:1992ts}
A.A.~Starobinsky,
JETP Lett.\  {\bf 55} (1992) 489
[Pisma Zh.\ Eksp.\ Teor.\ Fiz.\  {\bf 55} (1992) 477].

\bibitem{ace}
J.A.~Adams, B.~Cresswell and R.~Easther,
Phys.\ Rev.\ D {\bf 64} (2001) 123514
[arXiv:astro-ph/0102236].

\bibitem{Peiris:2003ff}
H.V.~Peiris {\it et al.},
Astrophys.\ J.\ Suppl.\  {\bf 148} (2003) 213
[arXiv:astro-ph/0302225].

\bibitem{Martin:2004yi}
J.~Martin and C.~Ringeval,
JCAP {\bf 0501} (2005) 007
[arXiv:hep-ph/0405249].

\bibitem{Kawasaki:2004pi}
M.~Kawasaki, F.~Takahashi and T.~Takahashi,
Phys.\ Lett.\ B {\bf 605} (2005) 223
[arXiv:astro-ph/0407631].

\bibitem{ars97}
J.A.~Adams, G.G.~Ross and S.~Sarkar,
Phys.~Lett.~{\bf B391} (1997) 271
[arXiv:hep-ph/9608336] and  
Nucl.~Phys.~{\bf B503} (1997) 405
[arXiv:hep-ph/9704286].

\bibitem{mukh98}
V.F.~Mukhanov and P.~J.~Steinhardt 
Phys.~Lett.~{\bf B422} (1998) 52-60
[arXiv:astro-ph/9710038].

\bibitem{star01}
A.A.~Starobinsky, S.~Tsujikawa and J.~Yokoyama 
Nucl.~Phys.~{\bf B610} (2001) 383-410
[arXiv:astro-ph/0107555].

\bibitem{groot01}
S.~Groot~Nibbelink and B.J.W.~van~Tent 
Class.~Quant.~Grav.~19 (2002) 613-640
[arXiv:hep-ph/0107272].

\bibitem{isobound}
M.~Beltran {\it et al.}, 
Phys.\ Rev.\ D {\bf 71} (2005) 063532
[arXiv:astro-ph/0501477].

\bibitem{hs04}
P.~Hunt and S.~Sarkar,
Phys.\ Rev.\ D {\bf 70} (2004) 103518
[arXiv:astro-ph/0408138].

\bibitem{fggklt01}
G.N.~Felder, J.~Garcia-Bellido, P.B.~Greene, L.~Kofman, 
A.D.~Linde, I.~Tkachev 
Phys.~Rev.~Lett. {\bf 87} (2001) 011601
[arXiv:hep-ph/0012142].

\bibitem{abc01-hyb}
T.~Asaka, W.~Buchm\"uller and L.~Covi
Phys.~Lett.~{\bf B510} (2001) 271-276
[arXiv:hep-ph/0104037].

\bibitem{cpr02}
E.J.~Copeland, S.~Pascoli and A.~Rajantie 
Phys.~Rev.~{\bf D65} (2002) 103517
[arXiv:hep-ph/0202031].

\bibitem{ggg02}
J.~Garcia-Bellido, M.~Garcia~Perez and A.~Gonzalez-Arroyo 
Phys.~Rev.~D67 (2003) 103501
[arXiv:hep-ph/0208228].

\bibitem{chmss2}
L.~Covi, J.~Hamann, A.~Melchiorri, A.~Slosar and I.~Sorbera,
work in progress.

\bibitem{Stewart:1993bc}
E.D.~Stewart and D.~H.~Lyth,
Phys.\ Lett.\ B {\bf 302} (1993) 171
[arXiv:gr-qc/9302019].

\bibitem{Lewis:2002ah}
A.~Lewis and S.~Bridle,
Phys.\ Rev.\ D {\bf 66} (2002) 103511 
(Available from \texttt{http://cosmologist.info}).

\bibitem{thx}
M.~Tegmark, A.J.S.~Hamilton and Y.Z.S.~Xu,
Mon.\ Not.\ Roy.\ Astron.\ Soc.\  {\bf 335} (2002) 887
[arXiv:astro-ph/0111575].

\bibitem{Cole:2005sx}
S.~Cole {\it et al.}  [The 2dFGRS Collaboration],
Mon.\ Not.\ Roy.\ Astron.\ Soc.\  {\bf 362} (2005) 505
[arXiv:astro-ph/0501174].

\bibitem{hst}
W.L.~Freedman {\it et al.}, Astrophys. J. {\bf 553} (2001) 47.

\bibitem{Shafieloo:2003gf}
A.~Shafieloo and T.~Souradeep,
Phys.\ Rev.\ D {\bf 70} (2004) 043523
[arXiv:astro-ph/0312174].

\bibitem{acbar}
J.~H.~Goldstein {\it et al.},
Astrophys.\ J.\  {\bf 599} (2003) 773
[arXiv:astro-ph/0212517].

\bibitem{boom}
C.J.~MacTavish {\it et al.},
arXiv:astro-ph/0507503.

\bibitem{cbi}
A.C.S.~Readhead {\it et al.},
Astrophys.\ J.\  {\bf 609} (2004) 498

\bibitem{maxi}
R.~Stompor {\it et al.}  [MAXIMA Collaboration],
arXiv:astro-ph/0309409.

\bibitem{vsa}
C.~Dickinson {\it et al.},
Mon.\ Not.\ Roy.\ Astron.\ Soc.\  {\bf 353} (2004) 732
[arXiv:astro-ph/0402498].

\bibitem{fields}
B.~Fields and S.~Sarkar,
arXiv:astro-ph/0601514.

\bibitem{Ashoorioon:2006wc}
  A.~Ashoorioon and A.~Krause,
  arXiv:hep-th/0607001.

\bibitem{eisenstein}
D.J.~Eisenstein {\it et al.},
Astrophys.\ J.\  {\bf 633} (2005) 560
[arXiv:astro-ph/0501171].

\bibitem{Wang:2002hf}
  X.~Wang, B.~Feng, M.~Li, X.L.~Chen and X.~Zhang,
  Int.\ J.\ Mod.\ Phys.\ D {\bf 14} (2005) 1347
  [arXiv:astro-ph/0209242].

\end{thebibliography}
\end{document}